\documentclass[aps,prb,showpacs,amsmath,superscriptaddress,reprint,floatfix,nobalancelastpage]{revtex4-1}

\usepackage[nottoc,numbib]{tocbibind}

\usepackage[colorlinks=true, citecolor=black, linkcolor=black, urlcolor=black]{hyperref}

\usepackage[all]{hypcap}
\usepackage{amsmath}
\usepackage{enumerate,bm}
\usepackage{graphicx}
\usepackage{grffile} % to handle dots in graphics filnames
\usepackage{color}
\usepackage{esint}
\usepackage{textpos}
\usepackage[usenames,dvipsnames]{xcolor}
\usepackage{subfigure}
\usepackage{flushend}
\usepackage{tabularx}
\usepackage{amssymb}
\usepackage{float}
\usepackage{subfigure}
\usepackage{ulem} %striketrhough
\newcommand{\qw}[1]{{#1}}

\usepackage{setspace}
\usepackage{verbatim}

\begin{document}
\title{First-principles modelling for time-resolved ARPES under different pump-probe conditions}
% First-principles description for time-resolved ARPES under different pump-probe conditions

\author{Umberto~De~Giovannini}
\email{umberto.degiovannini@gmail.com}
\affiliation{Max Planck Institute for the Structure and Dynamics of Matter and Center for Free Electron Laser Science, 22761 Hamburg, Germany}
\affiliation{IKERBASQUE, Basque Foundation for Science, E-48011, Bilbao, Spain, and Nano-Bio Spectroscopy Group, Departamento de Fisica de Materiales, Universidad del País Vasco UPV/EHU, 20018 San Sebastián, Spain}

\author{Shunsuke~A.~Sato}
\email{ssato@ccs.tsukuba.ac.jp}
\affiliation %[CCS]
{Center for Computational Sciences, University of Tsukuba, Tsukuba 305-8577, Japan}
\affiliation{Max Planck Institute for the Structure and Dynamics of Matter and Center for Free Electron Laser Science, 22761 Hamburg, Germany}

\author{Hannes~H\"ubener}
\email{hannes.huebener@gmail.com}
\affiliation{Max Planck Institute for the Structure and Dynamics of Matter and Center for Free Electron Laser Science, 22761 Hamburg, Germany}

\author{Angel~Rubio}
\email{angel.rubio@mpsd.mpg.de}
\affiliation{Max Planck Institute for the Structure and Dynamics of Matter and Center for Free Electron Laser Science, 22761 Hamburg, Germany}
\affiliation{Center for Computational Quantum Physics (CCQ), The Flatiron Institute, 162 Fifth avenue, New York NY 10010.}

\date{\today}

\begin{abstract}
    First-principles methods for time-resolved angular resolved photoelectron spectroscopy play a pivotal role in providing interpretation and microscopic understanding of the complex experimental data and in exploring novel observables or observation conditions that may be achieved in future experiments. Here we describe an efficient, reliable and scalable first-principles method for tr-ARPES 
    based on time-dependent density functional theory 
    including propagation and surface effects usually  discarded in the widely used many-body techniques based on computing the non-equilibrium spectral function
    and discuss its application to a variety of pump-probe conditions. We identify four conditions, depending on the length of the probe relative to the excitation in the materials on the one hand and on the overlap between pump and probe on the other hand. Within this paradigm different examples of observables of time-resolved ARPES are discussed in view of the newly developed and highly accurate time-resolved experimental spectroscopies.
    \end{abstract}

\maketitle

\section{Introduction}
Angular resolved photoelectron spectroscopy (ARPES) %\textcolor{red}{(SS: it seems that PES does not refer photoelectron spectroscopy but photo emission spectroscopy in the convention. But, I like more photoelectron than photoemission...)} 
is a long standing well established technique to observe the electronic energy dispersion in solids and surfaces. Indeed, it is the standard technique for the electronic structure of solids and the observed spectral functions are well understood from a theoretical point of view in terms of quasiparticles and their lifetimes, including manifestations of complex quasiparticle interactions~\cite{Damascelli.2003,Wang.2018}. Like many other spectroscopies, the possibility to perform time-resolved measurements has opened new avenues to observe dynamical processes and today, time-resolved (tr)-ARPES is a widely employed technique with a vast breadth of applications. This raises new challenges for the theoretical description of the observed dynamical effects as well as the simulation of the observation process itself.

Equilibrium ARPES is usually described within response theory, that captures the processes around the excitation and emission of the photoelectron. On the most basic level Fermi's Golden Rule, often referred to as the one step model~\cite{Hufner.2003usm,Damascelli.20044g}, describes the probability of producing a photo-electron by the dipole matrix element of the material state and the continuum. This can be refined by more sophisticated final states, modelling the detector etc. Such a description, however, neglects dynamical effects that occur in the photoemission process, such as the scattering of excited electrons \textit{before} they leave the material. This is attempted to be captured by the so called three step model, which accounts for the excitation, transport through the material, eventual emission and detection of the photoelectron, which allows to phenomenologically include the effects of excitation of or scattering with other material modes during the photo-hole creation process, such as plasmons, phonons as well as electron-electron scattering. The more rigorous description of many-body perturbation theory based on quantum field theory, avoids the separation in steps as well as phenomenological assumptions and has been very successful in describing the many-body interactions through self-energies and their signatures in the recorded spectral functions. However, for time-domain measurements, \qw{the equilibrium quasi-particle approximation is not sufficient,} because it does not account for the two-time nature of electronic correlations and the fundamental changes a system can undergo when it is excited\cite{Perfetto:2015il}.     

The formulation of the time-resolved ARPES process with non-equilibrium Green's functions\cite{Stefanucci2013} is considered as the most comprehensive theoretical description. Recently, it has been approached even as a first principles computational technique and demonstrated to be feasible for a real material\cite{sangalli2021excitons}, while lattice model based approaches have been implemented as well~\cite{Schler2020}. The non-equilibrium Greens functions approach allows to compute the time-dependent spectral function of materials, taking into account the complex many-body excitation\cite{Perfetto:2020wt},
de-excitation and decoherence\cite{Stefanucci:2021dl}, dynamical correlation\cite{Dendzik:2020df,Pavlyukh:2020ud}, \qw{excitations in strongly correlated systems\cite{Aoki.2014,Freericks.2009}} and lifetime effects. 

\begin{figure*}
  \centering
   \includegraphics[width=\textwidth]{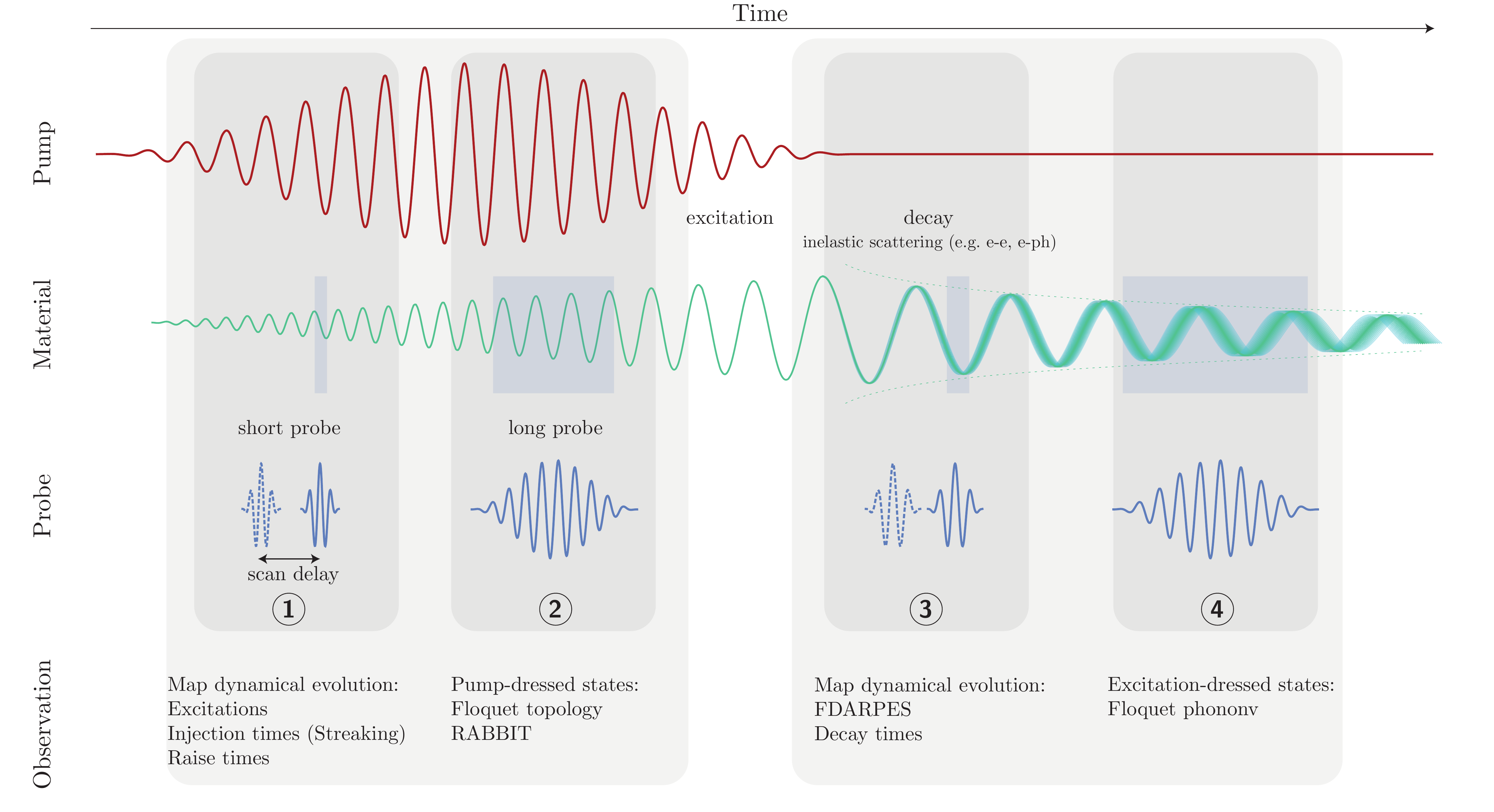}
   \caption{\label{fig:S1}
   {Different pump-probe conditions considered in this review. We distinguish the cases where pump (red line) and probe (blue lines) are either overlapping or not. The pump creates an excitation (green line) in the material. Depending on the relative time-scale between the probe \qw{envelope and the period of the excitation}, we further distinguish the cases of long and short probes. \qw{The blue shaded areas indicate the time scales at which the excitation is probed}. Each of these four conditions allow the observation of different phenomena. }
   }
\end{figure*}

In this article we will present another realistic and accurate approach for the first-principles computation of time-resolved ARPES that is based on time-dependent density functional theory (TDDFT)~\cite{DeGiovannini:2012hy,DeGiovannini.2013,Crawford-Uranga.2014,DeGiovannini:2017,Wopperer.2017,Krecinic.2018,Sato:2018jw, Hubener:2018id,DeGiovannini.2018, DeGiovannini.20187i1,deGiovannini:2020fv,Trabattoni.2020, Gatti.2020,Aeschlimann2021}. \qw{While in principle equivalent in scope and accuracy for coherent excitations to the quasiparticle method, TDDFT is limited by the available approximations to the complex exchange-correlation functional, but has the advantage of computational efficiency\cite{Onida.2002}}. Most applications of the Green's function methods give the quasiparticle spectral functions, whereas the method we discuss here explicitly computes the photo-ionization current and constructs the ARPES spectrum from its energy and momentum distribution, in a conceptual analogue to the photoelectron detector. TDDFT, being a theory for the electronic density, does not natively include electron-phonon coupling, however, it can be coupled to the classical ionic motion of the lattice through Ehrenfest molecular dynamics. This allows to describe the effects of coherent phonon excitations on the tr-ARPES spectrum~\cite{Hubener:2018id}. 
\qw{In such a quantum-classical coupled system it is not possible to accurately describe the exchange of energy between the lattice and electrons~\cite{Parandekar.2005,Parandekar.2006, Alonso.2020}, hence processes like thermalization and decoherence can not be described in this approach. Nevertheless}, decoherence effects can be accounted for in TDDFT by phenomenological scattering times and coupling to an density matrix~\cite{PhysRevB.99.224301}, while the coupling of quantum nuclear dynamics to a first-principles approach is under development~\cite{PhysRevLett.113.083003,PhysRevMaterials.3.023803,Abedi.2010,Abedi.2013}.

Here, we will discuss examples of observables of excited states in solids accessed via time-resolved ARPES and their theoretical description and first-principles simulation with TDDFT. To organise the examples we consider the paradigmatic regimes of short and long probes applied during and after the \textit{pump} pulse, as sketched in Fig.~1. The length of the probe pulse is here to be understood relative to the excited mode of the solid that is being created or the frequency of the pump that is applied. Probing a system with a long probe while the pump is applied allows the observation of Floquet states that have received considerable recent attention, because they offer the possibility to transiently alter the topology of a material \cite{PhysRevB.79.081406}. We will discuss how they are imprinted in the ARPES spectrum and issues and limitations that arise for they experimental detection. While Floquet states are usually associated with periodic driving fields, it is also possible to apply Floquet theory to states of a solid \textit{after} the pump, when a coherent eigenmode of the material is dressing the electronic structure. This possibility to create a field-free Floquet material, will be discussed alongside its signatures in ARPES spectra which can be observed under a long probe regime. By contrast, short probes allow to observe the electronic structure as it follows the oscillation of the excitation mode or pump field. In the latter case one can obtain streaking traces from solids, which gives access to the intrinsic times with which electrons are ejected into the continuum. In the time domain following the pump, the short probe regime allows to observe the coupling to the electronic structure to elemental modes, which we will show here for the example of electron-phonon coupling.    

It is worth mentioning that here we do not discuss the most widely considered observable of tr-ARPES, that is rise and decay times of excited state populations, because it is well covered elsewhere and in the other contributions to this special issue.

\section{Tr-ARPES with TDDFT \label{sec:tddft-arpes}}

TDDFT\cite{runge_density-functional_1984,Marques.2011,Ullrich.2012} provides a practical framework to treat quantum dynamics of many-electron systems out of equilibrium. The theory is based on the Runge-Gross theorem that establishes the one-to-one correspondence between the time dependent many-body wavefunction and the corresponding time-dependent density, $n(t)$. 
From this tenet, by employing  the Kohn-Sham (KS) scheme, one can reformulate the time-dependent problem of interacting electrons in terms of a fictitious non-interacting system -- the KS system -- having the same density as the interacting one at all times.  
In formulas this translates into a set of single particle equations for each, doubly occupied, KS orbital, $\varphi_{i,{\bf k}}({\bf r},t)$, representing $N$ electrons in the simulation cell, of the form 
\begin{multline}\label{eq:TDKS}
    i\frac{\partial}{\partial t}\varphi_{i,{\bf k}}({\bf r},t) =\Bigg[ \frac{1}{2}\left(\hat{\bf p}  +\frac{1}{c}\bf{A}(t)\right)^2+V_{ion}({\bf r})\\
    +V_H[n]({\bf r})+V_{xc}[n]({\bf r}) \Bigg] \varphi_{i,{\bf k}}({\bf r},t)
\end{multline}
with 
\begin{equation}
    n({\bf r},t)=  \sum_{\bf k}^{\rm BZ}\sum_{i=1}^{N/2} |\varphi_{i,{\bf k}}({\bf r},t)|^2,
\end{equation}
where, $V_{ion}({\bf r})$ indicates the external potential generated by ions, $V_H[n]({\bf r})$ is the Hartree potential, the electrostatic potential generated by the charge density of the electrons, $V_{xc}[n]({\bf r})$ is the exchange and correlation potential accounting for the many-body interaction, and the square bracket is a customary notation indicating the functional dependence on the density.
For convenience we consider the integrals in reciprocal space discretized over a grid of k-points covering the Brillouin zone.
The right hand side of Eq.~(\ref{eq:TDKS}) defines the KS Hamiltonian which we denote with $\hat{H}_{\rm KS}$. 
The coupling with external electric fields is expressed in the velocity gauge and in the dipole approximation, i.e. with a spatially uniform vector potential ${\bf A}(t)=-c \int_{-\infty}^t d\tau\,{E}(\tau)$. Here, we are concerned only with a purely electric vector potential. To include magnetic fields, one would have to employ the generalization of TDDFT to current-density functional theory. Although, here we use the dipole approximation, the scheme we discuss is generally valid and can be applied beyond the dipole approximation and deal with arbitrary fields. Atomic units, $\hbar=m_e=e=1$, are used trough the paper unless specified otherwise.

The time-dependent vector potential in Eq.~(\ref{eq:TDKS}) can represent any linear combination of fields and naturally includes the case of two fields, pump and probe, needed to perform tr-ARPES measurements. It must be noted that a perturbative treatment of the problem becomes unfeasible to describe the state-of-the-art experimental condition, where highly-nonlinear dynamics is induced by strong fields.
Direct integration approaches to Eq.~(\ref{eq:TDKS}), like the propagation in real-time of the equations, remain the only practical option.

\begin{figure}
  \centering
   \includegraphics[width=\columnwidth]{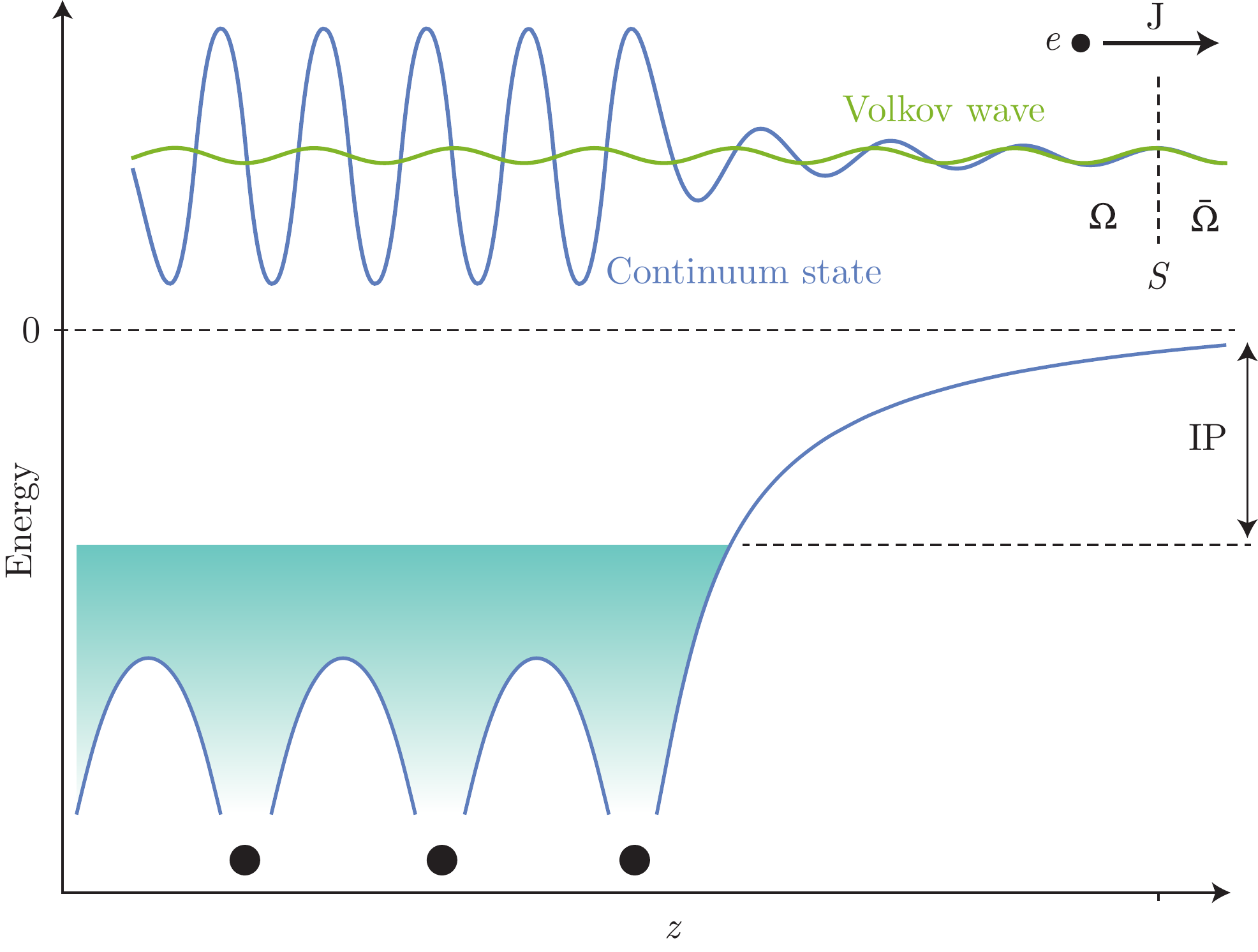}
   \caption{\label{fig:conitnuum}
   Schematic representation of the continuum states used to represent the photoelectrons away from the material. The assumption is made that beyond a certain distance from the materials surface the states are well represented by Volkov states.   
   }
\end{figure}

There are few options to extract the directly recorded tr-ARPES observable from a real-time TDDFT calculation which contains more than the electronic spectral function that is usually the result of other many-body approaches. 
The first option is to extract the photoelectron angular distribution from the flux of the ionization current through a surface with the t-SURFF method~\cite{Tao:2012ev, Wopperer.2017,DeGiovannini:2017}.
This approach requires to account explicitly for the surface termination of the material and to partition the spatial dimension perpendicular to the surface, $z$, into an inner, $\Omega$, and an outer, $\bar{\Omega}$ region like depicted in Fig.~\ref{fig:conitnuum}. 
This partitioning sets the stage for the approximation of the physical division between the material surface and the photoelectron detector that constitutes the basic ansatz of the approach.
In this division electrons in the inner region are fully interacting among each other, with the ions of the crystal and with the external field  while in the outer region they interact only with the external field. Formally it means that the Hamiltonian of the system is decomposed into the KS Hamiltonian in $\Omega$ and the Volkov Hamiltonian in $\bar{\Omega}$ as following

\begin{equation}\label{eq:tsurffham}
\hat{H}({\bf r},t)=\left\{
    \begin{aligned}
     \hat{H}_{\rm KS}({\bf r},t) &\text{ if } {\bf r} \in \Omega \\
     \hat{H}_{\rm V}({\bf r},t)  &\text{ if } {\bf r} \in \bar{\Omega}
    \end{aligned}
\right.   \,,
\end{equation}
where the Volkov Hamiltonian, $\hat{H}_{\rm V}$, is the Hamiltonian governing the dynamics of free electrons driven by an external homogeneous vector potential ${\bf A(t)}$, and is defined as
\begin{equation}
\hat{H}_{\rm V}({\bf r},t) = \frac{1}{2}\left( \hat{\bf p} + \frac{{\bf A}(t)}{c} \right)^2\,.
\end{equation}

The time-dependent Schr\"odinger equation (TDSE) associated with the Volkov Hamiltonian is solved by wavefunctions of the form
\begin{equation}\label{eq:volkowchi}
\chi_{\bf p}({\bf r},t) = \frac{1}{(2\pi)^{\frac{3}{2}}} e^{i {\bf p}\cdot{\bf r}} e^{-i \Phi({\bf p},t)}\,,
\end{equation}
with    
\begin{equation}
\Phi({\bf p},t)=\int_0^t {\rm d}\tau \frac{1}{2} \left( {\bf p} + \frac{{{\bf A}}(\tau)}{c} \right)^2\, ,
\end{equation}
that are eigenstates of the momentum operator and form a complete set.
This property allows one to use them as detector states and extract the momentum dependence of the KS orbitals in the continuum by projection
\begin{equation}\label{eq:phitochi}
\varphi_{i,{\bf k}}({\bf r},t) = \int {\rm d}{\bf p}\, b_{i,{\bf k}}({\bf p},t) \chi_{\bf p}({\bf r},t)\,.
\end{equation}
The total momentum probability, measured in the experiments, is simply obtained by summing up the 
squared modulus of the amplitude, $b_{i,{\bf k}}({\bf p},t)$, over states and k-points,
\begin{equation}\label{eq:tsurfPp}
\mathcal{P}({\bf p},t) =  \frac{2}{N} \sum_{\bf k}^{\rm BZ} \sum_{i=1}^{N/2}  \vert b_{i,{\bf k}}({\bf p},t)\vert^2\,.
\end{equation}

Using the continuity equation for the ionization current is then possible to formulate an equation for the amplitude~\cite{Wopperer.2017,DeGiovannini:2017} as a flux integral over the surface, $S$, separating $\Omega$ and $\hat{\Omega}$ as, 

\begin{equation}\label{eq:bpt}
 b_{i,{\bf k}}({\bf p},t) = -\int_0^t {\rm d }\tau \oint_S {\rm d } {\bf s}\cdot 
\chi^*_{\bf p}({\bf r}, \tau)    \left \{ \hat{\bf p}  + \frac{{\bf A}(\tau)}{c} \right \}\varphi_{i,{\bf k}}({\bf r},\tau)\,.
\end{equation}

This method provides a good estimate of ARPES matrix elements~\cite{DeGiovannini:2017}. This is due to the fact that the approximation involved in the Hamiltonian ansatz of Eq.~(\ref{eq:tsurffham}) requires that the continuum states of the KS Hamiltonian are well approximated by Volkov waves in the detector region $\bar{\Omega}$. From Fig.~\ref{fig:conitnuum} one can see that this approximation is mostly affected by the tail of the surface potential extending into the vacuum and that the error can be reduced simply by moving the surface $S$ further into the vacuum.
In practical calculations the dimension of the simulation box in $z$ is determined by the position of $S$ and by the absorbing boundary employed to prevent spurious reflections from the boundaries and therefore correctly model the vacuum region. The spectrum of possible boundaries conditions cover a wide range of alternatives going from simple complex absorbing potentials~\cite{DeGiovannini.2015} up to exterior complex scaling~\cite{Scrinzi.2010} and sophisticated open boundary conditions~\cite{Kaye.2020}.

A characteristic aspect of the semi-periodic boundary conditions employed in this method is that the surface of the material must be modeled by a slab geometry. This offers the possibility to study the electrons transport dynamics explicitly naturally beyond the \qw{sudden approximation}. Even though photoemission essentially probe surface properties because of the small mean free path of the photoelectrons in practical calculations one has to converge with respect to the number of layers needed to reach the bulk limit with clear implications in terms of computational costs. 
\qw{Still, many realistic applications can be treated by this method~\cite{deGiovannini:2016cb,DeGiovannini:2017,Schuler.2020,Gatti.2020,Aeschlimann2021}.}

In cases where the demand for accuracy is less stringent one can employ a simpler approach that requires only to model the dynamics in the bulk without the surface. 
This is the case for a method that can be derived by invoking the strong field approximation~\cite{Brabec.2000} where only the time evolution under the pump field is required explicitly and the effect of the probe is considered at the level of the one-step model. 
We can call this approach the time-dependent one-step model.
In this approximation the ionization amplitude for the $i$-th band can be  written~\cite{Sato_2020,DeGiovannini:2020chm} as
the time integral of the transition amplitude between an initial time-dependent state (pump driven), and a scattering state approximated by a Volkov wave resulting from the action of the probe field, ${\bf A}_{pr}(t)=\bm{\epsilon} F(t) $. In a formula 
\begin{equation}
    b_{i,{\bf k}}({\bf p}) = -i \int_\infty dt\, \langle \chi_{\bf p}(t)|\hat{\bf p} \cdot {\bf A}_{pr}(t) |\varphi_{i,{\bf k}}(t)\rangle\, ,
\end{equation}
where the bra-ket notation indicates spatial integration over the unit cell. 
This formula can be further simplified by using the fact that Volkov waves are eigenstates of the momentum operator and therefore  $\langle \chi_{\bf p}(t)|\hat{\bf p}\cdot {\bf A}_{pr}(t)|\varphi_{i,{\bf k}}(t)\rangle= {\bf p}\cdot {\bf A}_{pr}(t) \langle \chi_{\bf p}(t)|\varphi_{i,{\bf k}}(t)\rangle$. The ionization amplitude then becomes,
\begin{equation}\label{eq:tdFG}
    b_{i,{\bf k}}({\bf p}) = -i {\bf p}\cdot \bm{\epsilon} \sum_j \langle {\bf p } | \varphi_{j,{\bf k}}\rangle \int_\infty dt\, e^{i\Phi({\bf p},t)} F(t) c_{j,i,{\bf k}}(t)  
\end{equation}
\qw{where the time-dependent KS orbital is expanded in the basis of the KS orbital at equilibrium ($t=0$), $|\varphi_{i,{\bf k}}(t)\rangle = \sum_j c_{j,i,{\bf k}}(t) |\varphi_{i,{\bf k}}(t=0)\rangle$ with coefficients $c_{j,i,{\bf k}}(t)=\langle \varphi_{j,{\bf k}}|\varphi_{i,{\bf k}}(t)\rangle$}.
Note that Eq.~(\ref{eq:tdFG}) reduce to Fermi's golden rule in the limit where there is no pump and the probe is monochromatic, according to the one-step theory of photoemission.

The advantage of this approach is immediately apparent from Eq.~(\ref{eq:tdFG}) since it only requires the information of the time dependent projections from a dynamical evolution of the KS orbitals. 
This is an easily accessible quantity that can be available to any code capable to simulate the real-time dynamics of a solid. 

A limitation of this approach is that the photoemission matrix elements, $M=\langle {\bf p } | \varphi_{j,{\bf k}}\rangle $, are calculated as an expansion over plane waves, i.e. as a spatial Fourier transforms of the KS orbitals. 
While this approach is certainly convenient from the numerical stand point it delivers only qualitative results.
This is because plane waves are low-level approximations for the continuum states of the solid because of the presence of crystal fields, as one can appreciate by looking at Fig~\ref{fig:conitnuum}.
In order to improve the quality one should use more accurate continuum  states of the material such as time-reversed low-energy electron diffraction (LEED)~\cite{Hufner.2003usm} states that can be obtained with separate, static, calculations~\cite{PENDRY.1978}. 

\qw{It must be noted that basis expansion and final state approximations, like done by taking Eq.~(\ref{eq:tdFG}) in first-order perturbation theory, can cause an artificial gauge dependence that has to be carefully assessed in order to attain predictive results from theory~\cite{Foreman.2002, Schuler.2021abc}. 
Even though Eq.~(\ref{eq:tdFG}) and Eq.~(\ref{eq:bpt}) are defined using gauge-invariant physical states, the Volkov states, theoretical predictions based on Eq.~(\ref{eq:bpt}) are more robust because they are defined in terms of the gauge-invariant current operator instead of the dipole operator. Moreover the spectra obtained with t-SURFF can be systematically converged by moving the the flux surface $S$ further into the vacuum to improve the match between plane waves and continuum states whereas the overlap integral in Eq.~(\ref{eq:tdFG}) can only be improved, as discussed above, by changing the final state.
Indeed t-SURFF  has been widely used in the strong field community~\cite{Tao:2012ev,Tulsky.20204o} where gauge independence is crucial and applied, for instance, to laser-induced electron diffraction~\cite{Krecinic.2018,Trabattoni.2020}.}

\section{Overlapping pump and probe}
Here, we discuss the overlapping pump-prove regime, where pump and probe pulses are (partially) overlapped. In this regime, it is on the one hand possible to observe ultrafast excitation processes directly at their creation with short (sub-cycle) probe pulses and on the other hand create dressed states in the solids that emerge when the probe is longer than the pump carriers.

\subsection{Condition 1: Short-probe in the overlapping regime}
Experiments where the probe pulse extension in time is of the order or smaller than a single cycle of the pump carrier, c.f. Fig.~1, provide access to the sub-cycle excitation dynamics in a material.
Of particular interest for excitations in solids is the attosecond time scale (1 as = $10^{-18}$ s)  which is the natural scale of electron-hole dynamics.
Crucial for accessing this time scale is the ability to generate isolated attosecond pulses~\cite{Sansone.2006} which also opens the door on the vast field of the research on attosecond physics~\cite{Krausz.2009}.

In this regime tr-ARPES has been employed mainly for attosecond streaking~\cite{Itatani.2002}. Attosecond streaking is a pump-probe technique in which an XUV pulse creates a photoelecton wavepacket that is subsequently streaked by a moderately strong and carrier-envelope phase (CEP) stabilized IR field (typically of the order $I_{IR}\approx 10^{11}$~W/cm$^2$) and by this means encode a time information in the velocity spectrum of the photoelectrons. By carefully analysing the spectrogram obtained from the expansion of the photoelectron momentum distribution as a function of the time-delay it is possible to recover various contributions. Most notably the Eisenbud-Wigner-Smith (EWS) time delay of photoemission which defines the time that is takes for electrons to be released into the continuum during an ionization process. 

The working principle~\cite{Kienberger.2004wuc} is based on the fact that the electron momentum at the detector, ${\bf p}_f$, is constituted by the momentum with which the electron reaches the continuum, ${\bf p}_0$, shifted by the value of the pump field at the precise moment in time in which it get released, $t_0$, as
\begin{equation}
    {\bf p}_f={\bf p}_0-{\bf A}_{IR}(t_0)/c\,.
\end{equation}
The release time, $t_0$, is the observable that is directly measured albeit, in many cases, only as a quantity relative to other ionization channels. 
It can be thought as composed of the pump-probe delay, $\tau$, plus a generic time shift, $t_S$: $t_0=\tau+t_S$. The time shift is composed of a number of contributions originating from different sources such as atto-chirp, Coulomb tail or lattice scattering, etc. depending on the details of the experiment including the EWS delay.  

Attosecond streaking is considered one of the cornerstone of attosecond chronoscopy~\cite{Pazourek.2015,Seiffert.2017} with ample breadth of application that ranges from the first experiment on neon atoms~\cite{Schultze.2010} to molecules~\cite{Vos.2018, Biswas.2020} and clusters~\cite{Seiffert.2017}.
In solids it was first employed on Tungsten where $t_S$ $\approx 100$~as delay in the emission between localized core states and conduction band states  was observed~\cite{Cavalieri.2007fcl,Ossiander.2018}.
In Magnesium it was employed to investigate the screening in real-time~\cite{Neppl.2012,Neppl.2015} and to uncover the dependence on the initial state angular momentum in  WSe$_2$ van der Waals crystals~\cite{Siek.2017}.

\subsection{Condition 2: Long-probe in the overlapping regime}\label{sec:floquet}
When the probe is overlapping in time with a considerable portion of the pump it is possible to create a situation where resulting ARPES spectrum shows a quasiparticle spectrum that is not only determined by the intrinsic properties of the solid, but shows the spectrum of the pump-dressed solid. This has gained some recent attention, because such dressed states can have different properties, especially in terms of topology. In the limit of a constant pump envelope, such states, as well as their ARPES spectrum, can be described by Floquet theory\cite{Sambe:1973hi,Sentef:2015jp}. 

The Floquet theorem states that a time-dependent Schr\"odinger equation $i\frac{d}{dt}\phi^F_{\alpha}(t)=H(t)\phi^F_{\alpha}(t)$ with a time-periodic Hamiltonian ($H(t)=H(t+T)$), can be satisfied with the following form of the wavefunction, the Floquet states,
\begin{equation}\label{eq:floquet_states}
    \phi_\alpha^F(t) = e^{-iE_\alpha t}u_\alpha(t),
\end{equation}
where $u_\alpha(t)$ are the time-periodic function as $u_\alpha \left (t+T\right )=u_\alpha(t)$. Hence, one can expand the periodic part of the Floquet states as
\begin{equation}\label{eq:floquet-expansion}
u_\alpha(t)=\sum_{m=-\infty}^{\infty}e^{im\Omega t}u^\alpha_m,
\end{equation}
where $\Omega$ is the fundamental frequency $\Omega=2\pi/T$ of the applied laser.
Furthermore, inserting Eqs.~(\ref{eq:floquet_states}) and (\ref{eq:floquet-expansion}) into the original Schr\"odinger euation, one can obtain the following eigenvalue equation
\begin{equation}
    E_\alpha u^\alpha_n = \sum_m \mathcal{H}_{nm} u^\alpha_m, 
\end{equation}
where the Floquet Hamiltonian $\mathcal{H}_{nm}$ is given by
\begin{equation}\label{eq:HF}
    \mathcal{H}_{nm} = \frac{1}{T}\int_T dt H(t)e^{i(m-n)\Omega t} +m\Omega\delta_{mn} .
\end{equation}
Thus Floquet theory maps the problem of solving a differential equation in the time into a static eigenvalue equation in the energy domain, by means of the discrete Fourier expansion of Eq.~(\ref{eq:HF}). Any solution of the time-dependent Schr\"odinger equation can then be constructed as a linear combination of the Floquet states as $\phi(t) = \sum_\alpha f_\alpha \phi^F_\alpha(t)$, where the coefficients $f_\alpha$ depend on the boundary conditions. 

This expansion of the a driven quantum mechanical system is relevant for tr-ARPES in two ways: (i) because under these driving conditions ARPES records the spectrum of the Floquet Hamiltonian, $E_\alpha$ instead of the bare band energies\cite{Sentef:2015jp,PhysRevB.98.035138} and (ii) the intensity of the ARPES spectrum depends on the expansion coefficients $f_\alpha$. By analysing the photoemission matrix element, c.f. Eq.~(\ref{eq:fermi}), in terms of the Floquet expansion the photoemission probability amplitude reads\cite{DeGiovannini:2020chm}
\begin{align}\label{PESSFA}
    P({\bf p})=&\sum_\alpha \sum_{n} |f_\alpha|^2 
    |u_\alpha^{n}({\bf p})|^2 ({\bf p}\cdot {\bf A}^0_{pr})^2 \nonumber\\ &\delta({\bf p}^2/2 -E_\alpha +n\Omega-\omega).
\end{align}
This expression shows that the kinetic energy of the photoelectrons depends on the new, Floquet-eigenvalue $E_\alpha$ and moreover, that the spectrum shows this energy repeated at integer intervals of $n\Omega$. These are the Floquet sidebands, whose intensity depends on the norm of the harmonic component $u^\alpha_n$. The overall intensity of such a series of $E_\alpha$ sidebands is governed by the Floquet expansion coefficient $f_\alpha$, that depends on how much the dynamics of the system is described by the Floquet state $\psi^\alpha$. As such Eq.~(\ref{PESSFA}) not only shows how an ARPES spectrum of a continuously driven system can be analysed, it also illustrates that ARPES is a natural observable of a Floquet electronic structure~\cite{deGiovannini:2016cb}. This last point implies that many theoretical proposals relying on the formation of Floquet states, especially for topological systems, can, or rather should, be verified with ARPES experiments~\cite{schuler2020}.     

The observation of Floquet electronic structure, is however, not as straightforward as Eq.~(\ref{PESSFA}) suggests. In the above derivation, Floquet states are derived with an effective single-particle Hamiltonian without a significant contribution from other degrees of freedom. On the contrary, in real systems, an electron is not an independent particle but interacts with other degree of freedom such as phonons, defects and other electrons. As a result, the formation of single-particle Floquet states is disturbed by the surrounding environment, and signals from the Floquet bands are suppressed \cite{Schuler.2020c2h,Sato_2020,Aeschlimann2021}. From the viewpoint of many-body systems, in principle, one needs to consider not only electrons but also their surrounding environment in the framework of the Floquet theory in order to accurately describe dressed matter. If a single-particle degree of freedom is sufficiently isolated from the other degrees of freedom, a single-particle Floquet feature can be observed in real experiments.
Indeed, this mechanism has been observed in experiments on BiSe2~\cite{Wang:2013fe,Mahmood:2016bu} as an avoided crossing in ARPES sidebands.

So far, we discussed how electronic states in matter are dressed by light and how their properties are recorded in the tr-ARPES with the Floquet theory. These Floquet states are the initial states of the photoemission process induced via the probe pulse, and the corresponding final states (detector states) of the tr-ARPES simulation are the Volkov states as described in Sec.~\ref{sec:tddft-arpes}. Importantly, under a continuous-wave driving, these Volkov states can be also seen as the corresponding Floquet states. As a result, the observed tr-ARPES signals may contain additional side-band contributions from the dressed free-electron states; Volkov states. This effect is known as laser-assisted photoemission effect (LAPE)~\cite{Park.2014, Mahmood:2016bu}. This side-band formation due to the photo-dressed free-electron states play an essential role in the reconstruction of attosecond beating by interference of two-photon transitions (RABBIT) technique~\cite{Paul.2001,Klunder.2011}.
RABBIT is a technique that allows to recover the phase of the ionized electron wavepacket and the delay in photoemission. It can be view as technique complementary to streaking.
In solid is was applied to Nickel~\cite{Tao2016} and Tungsten~\cite{Heinrich.2021} crystals.

\section{Non-Overlapping pump and probe}
Here, we discuss a regime, where pump and probe pulses are sufficiently separated in the time-domain. In this regime one probes only the excitation created by the pump, rather than its build up or driven states. Depending on the relative time-scales of time-resolution and decay time one can either observe how the excitation is dissipated through the solid or one can measure dynamics of the excitation itself.

\subsection{Condition 3: Short-probe in the non-overlapping regime}
In this regime tr-ARPES can be used to probe excitations and eigenmodes of solids on their characteristic time scales, provided the delay resolution is fast enough to resolve the internal dynamics. In the following we will focus on the case where the pump has excited a coherent phonon and will analyse how in this condition tr-ARPES can give access to electron phonon coupling.

\begin{figure*}
  \centering
   \includegraphics[width=\textwidth]{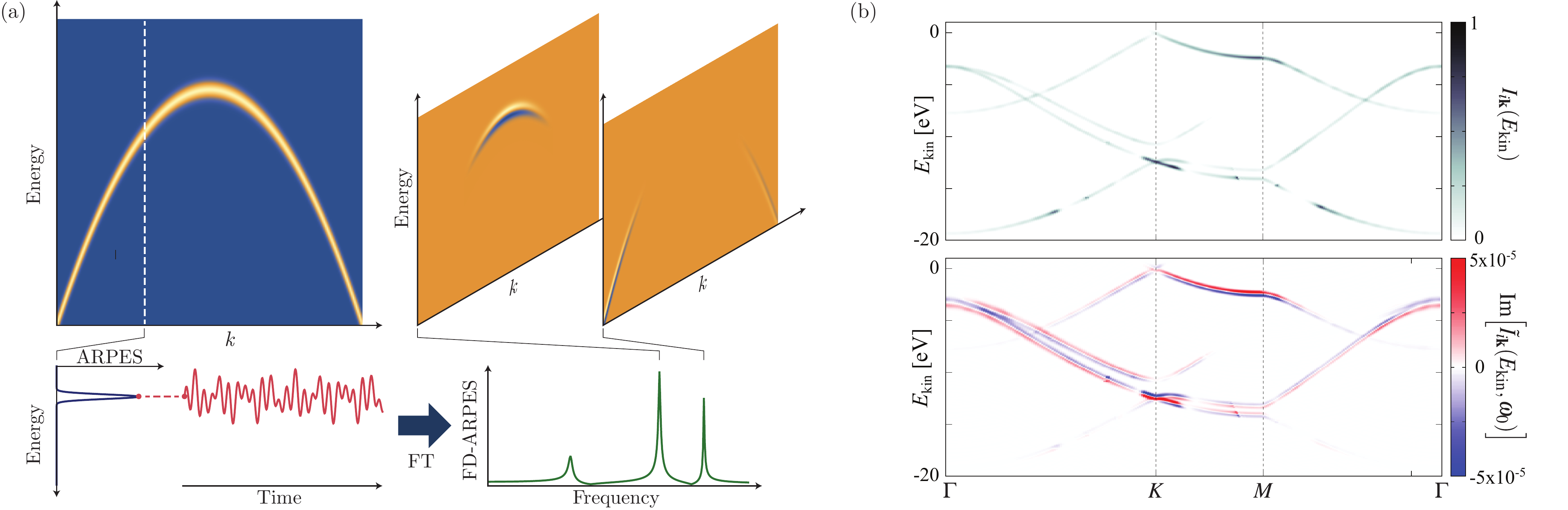}
   \caption{\label{fig:fdarpes}
   (a) Illustration of the FD-ARPES method. Under non-equilibrium conditions each point in an ARPES map acquires a time dependence that can be analyzed by Fourier transform. Selecting a particular Fourier component for all points allows the construction of the FD-ARPES map, that gives and energy and momentum resolved view of the couplign of the electronic structure to a mode with that particular frequency.
   (b) Computed equilibirum (upper panel) and FD-ARPES (lower panel) for Graphene the frequency of the E$_2g$ phonon mode, which has been coherently excited (adapted from~\cite{deGiovannini:2020fv}).
   }
\end{figure*}

The lattice displacement associated with a coherent phonon induces, to first order, a shift in the electronic band energies depending on the the strength of the electron-phonon coupling. This can be observed in tr-ARPES as an oscillation of the spectral function at a given $k$-point and by fitting this movement in energy on can extract the electron phonon coupling\cite{Rettig:2015ji,Gerber:2017bm}. Alternatively one can perform a Fourier transform of the time-series obtained by tr-ARPES for each point in the energy-momentum map and reconstruct the ARPES map at a given frequency\cite{hein:2019,Suzuki:2021fo,deGiovannini:2020fv}, c.f. Fig.~\ref{fig:fdarpes}(a). This latter option, Frequency-domain (FD)-ARPES, does not only provide a novel way of presenting the large amount of data collected in tr-ARPES measurements with novel high-resolution detectors, but it also presents the opportunity of processing the data differently, as we will discuss now.

To analyze the FD-ARPES signal based on the microscopic theory, we start from the Fermi's Golden Rule expression for the photoemission intensity at a given $k$-point and originating from a given band orbital $\psi$ with energy $\epsilon$
\begin{equation}\label{eq:fermi}
I_{i{\bf k}}(E_{\rm kin}) = |\langle f_{\bf p} | {\bf A}\cdot \hat{\bf p} |\psi_{i{\bf k}}\rangle|^2 F(\Omega+\epsilon_{i\mathbf{k}}-E_{\rm kin})
\end{equation}
where $E_{\rm kin} =p^2/2$ is the kinetic energy of the photoelectron, $f_{\bf p}$ is a final state with momentum $\mathbf{p}$, $F$ is the spectral lineshape and $\Omega$ is the energy of the probe. \qw{We now make the assumption that the lattice excitation is a coherent phonon or contains a substantial coherent component, so that the ionic motion can be described by a single classical trajectory.} Assuming a sinusoidal for the displacement of the lattice due to the coherent phonon generation, $u(\tau)=u_0 \sin{\omega_0 \tau}$ with the maximum lattice displacement $u_0$ and the phonon frequency $\omega_0$, we can account for the effect of the \textit{adiabatic} coupling to coherent lattice motion by introducing a parametric dependence of the band energy and orbital on the lattice displacement $u$ through ordinary perturbation theory:
\begin{align}\label{eq:dE}
    \epsilon_{i\mathbf{k}}[ u(\tau)] & =  \epsilon_{i\mathbf{k}} + \Delta_{i{\bf k}} \sin{(\omega_0\tau)} \\
   | \psi_{i\mathbf{k}}[ u(\tau)]\rangle & = |\psi_{i\mathbf{k}}\rangle + |\delta \psi_{i\mathbf{k}} \rangle  \sin{(\omega_0\tau)} \label{eq:dpsi}
\end{align}
where $\Delta_{i{\bf k}} = u_0\langle \psi_{i\mathbf{k}} |\delta V |\psi_{i\mathbf{k}} \rangle$ is the change in orbital energy at the maximum displacement $u_0$ of the lattice, and $g_{ii}(\mathbf{k})=\langle \psi_{i\mathbf{k}} |\delta V |\psi_{i\mathbf{k}}\rangle$ is the diagonal element of the standard electron-phonon coupling matrix~\cite{Giustino:2017ge}, while the term $|\delta \psi_{i\mathbf{k}} \rangle = u_0 \sum_{i\neq j} \frac{g_{ij}({\bf k})}{ \epsilon_{i\mathbf{k}}- \epsilon_{j\mathbf{k}}} | \psi_{j\mathbf{k}}\rangle$ depends on the interband electron-phonon coupling. Inserting these expansions into the Fermi's Golden Rule expression, Eq.~(\ref{eq:fermi}), gives a model for the tr-ARPES intensity of an adiabatically coupled coherent phonon:
\begin{equation}
I_{i{\bf k}}(E_{\rm kin},\tau) = |\langle f_{\bf p} | {\bf A}\cdot \hat{\bf p} |\psi_{i{\bf k}}[u(\tau)]\rangle|^2 F(\Omega+\epsilon_{i\mathbf{k}}[u(\tau)]-E_{\rm kin})
\end{equation}
and the FD-ARPES map is obtained by Fourier transforming with respect to $\tau$. It is important to note, that despite the linear expansion of the band properties, this expression is clearly non-linear in the lattice displacement. Since the coupling is adiabatic, the Fourier transform can only have components at the frequency of the phonon and its harmonics. For the fundamental frequency this gives
\begin{align}\nonumber
    \tilde{I}_{i{\bf k}}(E_{\rm kin},\omega_0) =& \frac{1}{2\Delta} M  \tilde{J}_1\bigg(\frac{E}{\Delta}\bigg)* F(E) +\\
      &\frac{i}{2\Delta}  \delta M  \left[\tilde{J}_0\bigg(\frac{E}{\Delta}\bigg)+\frac{3}{2}\tilde{J}_2\bigg(\frac{E}{\Delta}\bigg)\right] * F(E) \label{eq:FD_full}
\end{align}
where $\tilde{J}_n$ are Fourier transforms of Bessel functions, $*$ denotes the convolution product, $\delta M=Re[\langle \psi_{i{\bf k}}|{\bf A}\cdot \hat{\bf p}|f_{\bf p}\rangle \langle f_{\bf p}|{\bf A}\cdot \hat{\bf p}| \delta\psi_{i{\bf k}}\rangle ]$ and we have used the shorthand $E=\Omega+\epsilon_{i\mathbf{k}}+E_{\rm kin}$. 

The expression in Eq.~(\ref{eq:FD_full}) gives the FD-ARPES map at the phonon-frequency for adiabatic electron-phonon coupling and an example for such a spectrum is given in  Fig.~\ref{fig:fdarpes}(b), showing the FD-APRES spectrum for Graphene at the frequency of the coherently excited E$_2g$ phonon mode. Generally, FD-ARPES spectra can be used to disentangle a large number of processes linked to the electron-phonon coupling that affect the ARPES spectrum. First of all, we note that the expression in Eq.~(\ref{eq:FD_full}) is purely imaginary, reflecting the assumption of the adiabatic electron dynamics and the form of the lattice displacement, $u(\tau)=u_0 \sin{\omega_0 \tau}$. The complex phase is determined by the phase of the time dependence in Eqs.~(\ref{eq:dE}-\ref{eq:dpsi}) and is therefore somewhat arbitrary. However, if the measurements detects another phonon mode at a different $\omega_0$ the relative phase of the FD-ARPEES maps can reveal different excitation mechanism for the two coherent modes, i.e. a displasive vs. an impulsive Raman process. However, maybe more telling can be the absence of a unique complex phase across an FD-ARPES map. In this case localised deviations of the complex phase directly signal regions in the Brillouin zone where the electron-phonon coupling is not adiabatic, pointing to underlying resonances and points where the electrons and phonons participate in other excitations. Secondly, the FD-ARPES expression of Eq.~(\ref{eq:FD_full}) provides a way to distinguish the type of electron phonon coupling. We note that the first term has odd symmetry as a function of energy and depends only on $\Delta$ while the second term even symmetry and contains $\delta M$. This means that if an FD-ARPES map is recorded that features lineshapes of purely odd symmetry, all the underlying electron-phonon coupling is provided by the diagonal coupling matrix elements $g_{ii}(\mathbf{k})$. Conversely, a deviation from the odd symmetry means that $|\delta \psi_{i\mathbf{k}}\rangle$ contribute to the FD-ARPES signal, meaning that band-orbitals are affected by the electron-phonon coupling and that off-diagonal coupling matrix elements $g_{ij}(\mathbf{k})$ are involved. Finally, we have shown in Ref.~\cite{deGiovannini:2020fv} that using a perturbative expansion in terms of small lattice distortions of small coupling, one can extract the diagonal matrix elements $g_{ii}(\mathbf{k})$ from FD-ARPES data, thus making the matrix element and observable of this spectroscopy.

We note that a necessary condition for FD-ARPES is that the underlying tr-ARPES measurement is able to resolve dynamics on the the time-scale of the phonon frequency. Hence, it falls in the short probe category, where short is relative to a previously excited mode. Next, we will turn to the opposite case, where the probe is \textit{longer} than the period of the excited mode. 

\subsection{Condition 4: Long-probe in the non-overlapping regime}
In this regime the probe itself does not resolve any dynamics on the time scale of the of the excitation. However, it neither results in a mere average of the signal. Instead, the correct picture to describe this regime is the Floquet expansion, despite the fact that there is no pump anymore. We will illustrate this paradigm again with the example of a coherent phonon excitation.

We noted section~\ref{sec:floquet} that when the probe is longer than the pump duration it may record the Floquet spectrum of a material, because the electronic levels appear as dressed by the photons to the probe. The same picture applies in the absence of a pump, when a coherent phonon has been excited in the system. In this case the Hamiltonian is periodic in time, $H(t)=H(t+T)$ where the periodicity now results from the oscillating lattice potential instead of an external field. 

\begin{figure}
  \centering
   \includegraphics[width=\columnwidth]{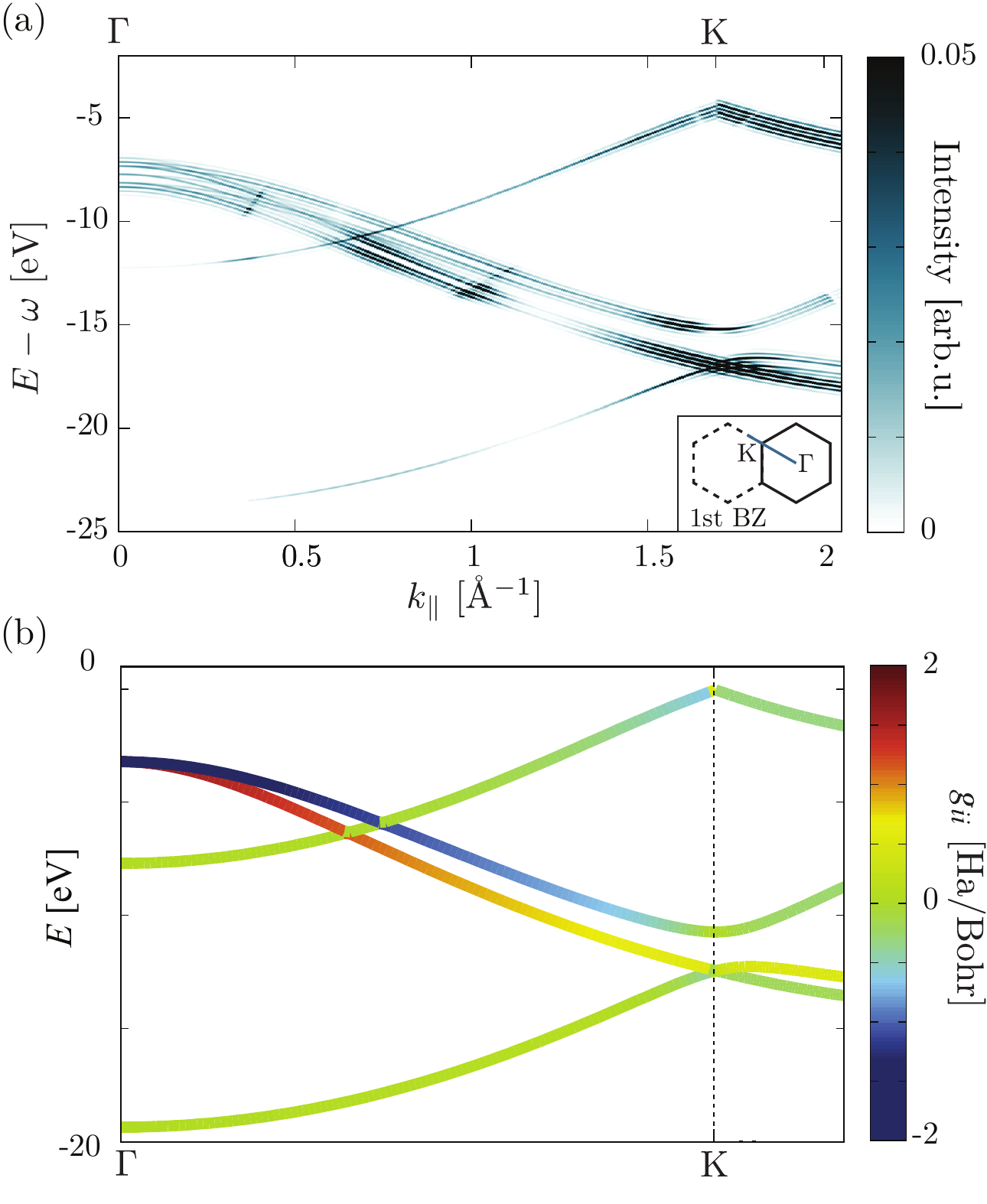}
   \caption{\label{fig:floquet_phonon}
   (a) Computed ARPES spectrum of Graphene with the E$_2g$ mode under Floquet-phonon conditions (adapted from~\cite{Hubener:2018id}). 
   (b) Electron-phonon matrix elements of the E$_2g$ mode of Graphene (adapted from~\cite{deGiovannini:2020fv}). Comparison between the panels shows that the number of visible sideband broadly tracks the strength of the electron-phonon coupling. Note, that the ARPES result also contains photoelectron matrix elements that suppress some bands more than others, leading to some deviations from this interpretation.
   }
\end{figure}

To see how the resulting Floquet features depend on the material properties it is instructive to write the time-dependent electronic Hamiltonian coupled to a coherent phonon mode $\omega_{\nu}$ at a given $k$-point as
\begin{equation}
    H_{\mathbf{k}\nu}(t) =  \epsilon_{n\mathbf{k}} c^\dagger_{n\mathbf{k}}c^\dagger_{n\mathbf{k}} + \sum_{mn} g_{mn\nu}(\mathbf{k}) c^\dagger_{m\mathbf{k}}c^\dagger_{n\mathbf{k}}\mathbf{u}_{\nu} \sin (\omega_{\nu} t).
\end{equation}
where $u_{\nu}$ is the amplitude of the coherent lattice motion along the phonon eigenvector and $g$ is again the electron-phonon matrix element. Optically excited coherent phonons are typically zone center modes, hence here we have additionally assumed that the phonon carry no moment, i.e. $q=0$. This formulation now shows that on the one hand the amplitude $u$ of the lattice motion takes the role of the laser intensity for the Floquet states and on the other hand the electron-phonon matrix elements $g$ that of the optical dipole matrix elements. This means that Floquet-phonon features in ARPES depend directly on the single band electron-phonon matrix element, which is somewhat analogous to the above example of the FD-ARPES case. Indeed, one can view an ARPES measurement of the Floquet-phonon case as the limit of the FD-ARPES case, where the probe time is very long, giving a static, rather than a time-resolved result. Fig.~\ref{fig:floquet_phonon} shows an example of a computed ARPES spectrum in the Floquet-phonon regime, where the momentum and energy dependence of the the sideband features are clearly follows that of the electron-phonon matrix elements. 

Finally, we comment on the similarity between Floquet satellites and other satellite or sideband features that are ubiquitous in many experimental ARPES spectra\cite{Damascelli.2003,RevModPhys.93.025006}. Strong coupling between the electronic structure and quasiparticle eigenmodes, such as plasmon or indeed phonons, can lead to sidebands in the ARPES of the equilibrium material, when the creation of the hole in the photoemission process also results in the excitation of a quasiparticle eigenmode of the solid and hence to a reduced energy of the photoelectron. While this phenomenology is very similar to that of the Floquet, it results from a very different physical situation: Floquet-ARPES probes an excited solid while the sidebands observed in equilibrium result from excitations due to the probing process itself. The two processes can be distinguished however, because in the equilibrium solid sidebands mostly occur below the main band, because the photoelectron has lost energy. In Floquet-ARPES by contrast, the already active eigenmode can supply the photoelectron with additional energy and hence sidebands occur symmetrically above and below the main peak.

\section{Outlook}
Here we have sketched an overview of the developments we have carried out over the last years to develop a reliable, efficient and widely applicable theoretical framework to address all kind of pump-probe conditions and their respective observables. This is by no means attempting at being comprehensive as many particular conditions and regimes have been omitted. A major point of interest in tr-ARPES measurements is the observation of dynamical correlation\cite{Gatti.2020}, quasiparticle dynamics, coupling (decoherence and dissipation effects) and dressing of quasiparticles. Furthermore, the discussion here has been largely concerned with single photon processes (with the exception of the condition 1, the standard Floquet case). However, in tr-ARPES the possibility to create two (pump+probe) or multiple photon pathways that can interfere and thus give a different kind of characteristic spectrum is also possible and widely employed. The theoretical tool described here, includes such processes out-of-the-box and remains to be explored. Another yet to be explored field is the systematic employment of the FD-ARPES method to eigenmode excitations other than phonons. For instance plasmon or magnon coupling should lend itself to be systematically characterised by this approach, as well as other collective modes such as the charge-density wave transitions or excitons in the exciton insulator phase. Furthermore, FD-ARPES might be employed to monitor the electron-phonon coupling under non-equilibrium conditions, thus possibly giving insight into phenomena such as light-induced superconductivity. 
One further prospect (related to streaking and RABBIT) could be to study how exciton are formed during and get further observations on the "birth of and exciton"~\cite{Voorhis2015} or light induced topological phases of matter, for instance by monitoring Weyl formation\cite{Hubener:2017ht} and dynamics, among others. Recent success in tr-ARPES experiment investigating the decay dynamics of excitons in TMD~\cite{Madeo.2020,Dong:2021do} suggests that pursuing this route could be feasible.

\section{Acknowledgment}
This work was supported by the European Research Council (ERC-2015-AdG694097), the Cluster of Excellence ’CUI: Advanced Imaging of Matter’ of the Deutsche Forschungsgemeinschaft (DFG) - EXC 2056 - project ID 390715994, Grupos Consolidados (IT1249-19), partially by the Federal Ministry of Education and Research Grant RouTe-13N14839, the DFG -- SFB-925 -- project 170620586, and JSPS KAKENHI Grants No. JP20K14382. The Flatiron Institute is a division of the Simons Foundation.

%\bibliographystyle{ieeetr}
% \bibliography{bibliography}


\end{document}